\documentclass[showpacs,showkeys,prl,amsmath,amssymb,floatfix,preprint]{revtex4}
\usepackage{graphicx}% Include figure files
\usepackage{dcolumn}% Align table columns on decimal point
\usepackage{bm}% bold math
\usepackage{natbib}
\usepackage{subfig}
\usepackage{epstopdf}
\usepackage{natbib}
\usepackage{amsmath} 

\begin{document}

\title{Universality in dynamic wetting dominated by contact line friction}
\author{Andreas Carlson}
 \email{andreaca@mech.kth.se}
\author{Gabriele Bellani}
  
\author{Gustav Amberg}

\affiliation{Department of Mechanics, Linn\'e Flow Center,\\
 The Royal Institute of Technology,\\
 Stockholm, Sweden.
}

\date{\today}

\begin{abstract}
We report experiments on the rapid contact line motion present in the early stages of capillary driven spreading of drops on dry solid substrates. The spreading data fails to follow a conventional viscous or inertial scaling. By integrating experiments and simulations, we quantify a contact line friction ($\mu_f$), which is seen to limit the speed of the rapid dynamic wetting. A scaling based on this contact line friction is shown to yield a universal curve for the evolution of the contact line radius as a function of time, for a range of fluid viscosities, drop sizes and surface wettabilities.
\end{abstract}

 
\maketitle

The interest in moving contact lines is increasing due to the need for design of fluid applications at small scales, since these often rely on manipulation or control of two-phase flow. Examples of such are microfluidic systems, sintering, printing, coating and immersion lithography techniques to name a few. Spontaneous spreading occurs in the deposition and formation of micron sized drops in biomedical applications and when rewetting the lubricating film covering the eye. 

A generic example of dynamic wetting is the spreading of a spherical liquid drop as it comes in contact with a dry solid surface. Its spreading after contact is dominated by different physical mechanisms at various stages in the temporal evolution. If the drop radius is less than its capillary length, the flow is mainly driven by the interfacial energy of the drop and the substrate surface energy. The contact line is formed at the intersection of the drop liquid-air interface and the solid substrate, where the dynamic contact angle is defined as the angle between the liquid-air interface and the substrate. For a moving contact line, the interface is typically distorted near the solid surface, giving rise to a free surface capillary force, which may pull the contact line forward. These forces are balanced by different rate-limiting processes, such as viscous dissipation \cite{huh1970} and inertia \cite{Bird:2008kk}, which all act to reduce the contact line speed. 

It is well known that the classical hydrodynamic theory predicts a divergence of viscous stress at the contact line. Therefore it might be expected that the spreading is dominated by the viscous dissipation in the bulk. By regularizing the viscous dissipation, a model for the spreading in viscously dominated wetting is established \cite{voinon1976a}. This is often referred to as Tanner's law where the spreading radius ($r$) evolves as $r \sim R(\frac{\sigma t}{\mu R})^{{\frac{1}{10}}}$, where $\sigma$ is the surface tension coefficient, $R$ the initial drop radius and $\mu$ the viscosity. This model, which holds promise if the drop evolves slowly and has a shape similar to a spherical cap, has explained many experiments. However there are many wetting phenomena that it does not describe, illustrating that there are other mechanisms influencing or dominating the spreading. 

One example is the spontaneous spreading of a water drop as it comes in contact with a low energy substrate. Experiments indicate here that the acceleration of liquid in the bulk of the drop is resisting contact line motion. An inertial spreading is found to follow $r\sim R\left(\frac{R^3\rho}{\sigma}\right)^{{\frac{1}{4}}}\cdot t^{\frac{1}{2}}$ \cite{Biance:2004la} ($\rho$ is the density), but by making the substrate more hydrophobic a different exponent for the spreading radius was found \cite{Bird:2008kk}. The hydrodynamic model cannot fully capture wetting at high capillary numbers (given by the ratio of the viscous and surface tension force) \cite{CHEN:1995fk}, and dynamic wetting experiments of viscous ($1$Pa s) drops \cite{Bliznyuk:2010fk}. In the latter case the spreading radius was observed to increase as the square root of time ($r\sim t^{1/2}$). 

%Hydrodynamic theory can not unify these seemingly similar wetting phenomena, and this suggests that there are additional mechanisms influencing the spreading. We recently, \cite{carlson:exp2010muf} extracted from simulations a dissipation contribution at the contact line that arises from a friction related to the motion of the contact line itself. A frictional related to the contact line was also described in macroscopic simulations by \cite{ren:2010PoF}. On the nanoscale, a force was also found at the contact line \cite{Ren:2007qc} and the observed, very rapid, spreading was suggested to be a diffusive or active process. Questions still linger about the actual molecular mechanism that moves the contact line, where both substrate heterogeneity \cite{Prevost:1999fk} and molecular hopping between potential wells (adsorption sites) \cite{blake1969} have been suggested as possible explanations.

De Gennes \cite{degennes:1985} postulated that there might be another non-hydrodynamic dissipative contribution arising from the contact line itself. This macroscopic dissipation was defined by a friction factor local at the contact line, which has the same units as viscosity. Others \cite{carlson:exp2010muf, Ren:2007qc, Prevost:1999fk, ren:2010PoF, blake1969} have also discussed the importance of local non-hydrodynamic effects at the contact line, with different interpretations of its microscopic origin. Recently \cite{duvivier2011} a friction factor was estimated from the molecular kinetic theory by fitting the experimental spreading radius for drops with different viscosity. These experimental observations are collected at much later time scales than presented here, and the value for this friction factor is an order of magnitude larger than our numerical measurements.

%We report here values for the contact line friction factor that are obtained by integrating experiments and  simulations based on the Cahn-Hilliard Navier Stokes equations \cite{Carlson:2009db}. 
By integrating experiments and axi-symmetric simulations based on the Cahn-Hilliard Navier Stokes equations \cite{Carlson:2009db, Carlson:2012epl} we estimate values for the friction factor ($\mu_f$) that appears in the free energy formulation. Theoretically, the friction factor generates a local dissipation at the contact line through its boundary condition. Here, particular attention is devoted to the very first stage of a spontaneous spreading process that is far from equilibrium. The experimental data cannot be rationalized as viscous or inertial effects. The data set collapses for a scaling law based on the numerically measured contact line friction parameter $\mu_f$, even for a wide range of viscosities (1-85mPa s), different drop sizes and surface energies. These results indicate that local dissipation at the contact line, interpreted as a contact line friction, is limiting spreading.

Both experiments and numerical axi-symmetric simulations of drop spreading have been performed. The simulations are based on the Cahn-Hilliard Navier-Stokes equations \cite{Carlson:2009db}. In terms of phenomenological thermodynamics one can postulate the free energy ($F$) for a binary fluid  $F=\int \left(\frac{\sigma}{\epsilon}\Psi(C) + \frac{\sigma \epsilon}{2} |\nabla C|^2\right) d\Omega +\int \left((\sigma_{sl}-\sigma_{sg})g(C)+\sigma_{sg}\right) d\Gamma$. The volumetric ($\Omega$) free energy consists of two terms representing the bulk ($\frac{\sigma}{\epsilon}\Psi(C)$) and interfacial energy ($ \frac{\sigma \epsilon}{2} |\nabla C|^2$), respectively. $\Psi=\frac{1}{4}(C^2 -1)^2$ is a double-well function with two minima, giving the equilibrium values of the order parameter $C$, as $C=-1$ for gas and $C=1$ liquid.
The diffuse interface width ($\epsilon$) is chosen to be the same as the spatial resolution in the experiments $\epsilon=7.5\mu m$. Important to note, however, is that in \cite{carlson:exp2010muf} $\epsilon$ has been varied one order of magnitude, without any noticeable change in the results or any increase in viscous dissipation.

 The surface energy of the wet substrate is $\sigma_{sl}$, and the dry ($\sigma_{sg}$). $g(C)=\frac{1}{4}(2+3C-C^3 )$ is chosen to give $g(1)=1$ and $g(-1)=0$, thus producing the corresponding wet or dry surface energy of the substrate. 

By making a variation in $F$ with respect to the concentration, one obtains an expression for the chemical potential ($\delta F/\delta C$). If accounting for the effects of convection of the concentration, that would equal the flux due to gradients of the chemical potential, the Cahn-Hilliard equation is recovered, which along with the Navier Stokes equations forms a theoretical basis for modeling of wetting \cite{Carlson:2009db} with a no-slip on the wall.

By retaining any perturbation in the concentration at the wall, a general wetting boundary condition for the concentration at the solid surface appears \cite{Jacqmin:2000xq},
\begin{equation}
\epsilon \mu_{f} \frac{\partial C}{\partial t}= -\epsilon \sigma \nabla C \cdot \mathbf{n} + \sigma \cos(\theta_e) g'(C).
\label{bc}
\end{equation}
We interpret here $\mu_{f}$ as a friction factor at the contact line. $\theta_e$ is the equilibrium contact angle.

Experiments of spontaneously spreading drops have been carried out through high-speed imaging (150kfps) for different viscosities and coatings (oxide, silane, teflon) on Si-wafers. The viscosity was changed by using different glycerin-water mixtures, of glycerin mass-fraction $0\%$, $50\%$, $65\%$, $72.5\%$, $82.5\%$ , corresponding to viscosities [1, 6.6, 14, 31, 85] mPa s, respectively. The different viscosities do not give any significant change in equilibrium contact angle ($\pm2^{\circ}$), which were measured as $\theta_e=[20^{\circ}, 60^{\circ}, 109^{\circ}]$ for oxide, silane and teflon coatings. 

The axi-symmetric Cahn-Hilliard Navier Stokes simulations mimic the experiments using the same material properties (density, viscosity, surface tension and equilibrium angle) as measured from experiments. To obtain the experimentally observed spreading behavior, an additional dissipation at the contact line was necessary through a non-zero $\mu_f$ \cite{Carlson:2009db}. $\mu_f$ was determined by obtaining a direct agreement between simulations and experiments, enabling a direct measurement of $\mu_f$ even in the presence of other contributions such as viscosity and inertia \cite{Carlson:2012epl}. The values for $\mu_f$ are reported in table I for all the surfaces and viscosities. A non-monotonicity in $\mu_f$ is observed for pure water for the $SiO_2$ and silane coating, the same dependency was reported in \cite{Carlson:2009db} when comparing with similar experiments \cite{Bird:2008kk}. We can at the present time not explain this non-monotonicity for pure water. Fig. 1 shows the excellent agreement between simulations and experiments for a water and a glycerin--water ${82.5\%}$ drop with an initial radius $R\approx0.5mm$. See also Supplemental Material at \cite{SM}. Fig.1a shows the initial condition in the experiments and simulations, and the field of view in the experiments (dashed box). The same window was extracted from the numerics, however the whole drop was simulated.
\begin{figure*}
\centering
\includegraphics[width=1.00\linewidth,angle=0]{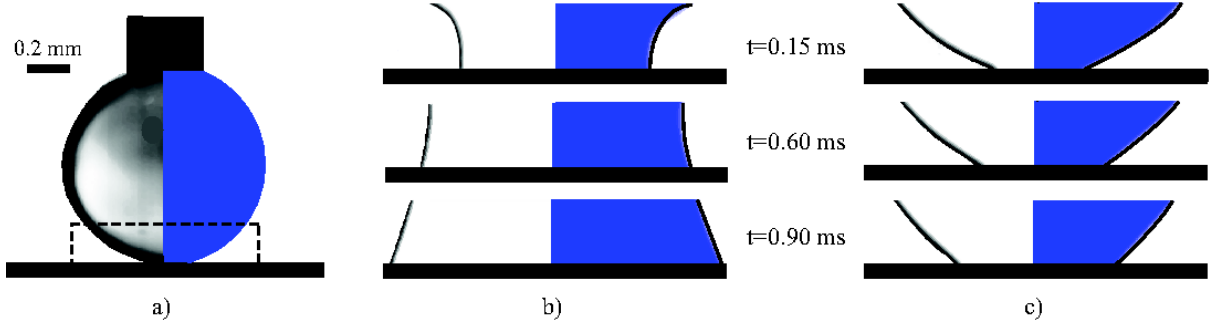}
\caption{Panel a) Illustrates the initial condition for the experiments and the numerical simulations, where a drop held at the tip of a needle is brought into contact with a dry solid substrate. The dashed box in the figure shows the field-of-view in the experiments. In panels b) and c) are shown the drop shape near the substrate, at times $t=0.15ms$, $t=0.60ms$ and $t=0.90ms$, after initial contact. Each panel shows a composite of experiment (left) and simulation (right). The black solid line in the right half that is plotted on top of the simulation result, illustrates experimental interface shape. b) A water drop spreading on an oxidized Si-wafer ($\theta_e=20^{\circ}$, viscosity $\mu_{H_2O}=1$mPa s). c) Glycerin ${82.5\%}$ drop spreading on an oxidized Si-wafer ($\theta_e\sim20^{\circ}$, viscosity $\mu_{glycerin_{82.5\%}}=85$mPa s).}
\label{fig:fricvsvisc}
\end{figure*}
%\caption{Each subfigure shows a direct comparison between experiment (left) and corresponding simulation (right) of a droplet with an initial radius $R=(0.5\pm 0.02) mm$ at $0.33ms$ after start of spreading. The mass fraction glycerin in water is given on the top and the three different substrates are indicated to the left.}

\begin{table}
\label{muf}
\begin{center}
\begin{tabular}{c|c|c|c|c|r}
\hline
\hline
Mass fraction glycerin   &$0\%$ & $50\%$ & $65\%$ & $72.5\%$ & $82.5\%$\\ \hline

${SiO_2}$ [Pa s]&0.15 & 0.33& 0.51 & 0.66& 1.02\\\hline
${Silane}$ [Pa s]&0.17 & 0.26& 0.33 & 0.41& 0.80\\\hline
${Teflon}$ [Pa s]&0.07 & 0.06 & 0.09 & 0.10 & 0.19\\\hline
\hline
\end{tabular}
\caption{Values for the contact line friction parameter $\mu_f$ [Pa s] for different viscosities and substrates ($SiO_2, Silane, Teflon$) measured from the numerics.}
\end{center}
\end{table}

\begin{figure}
{\includegraphics[width=0.80\linewidth]{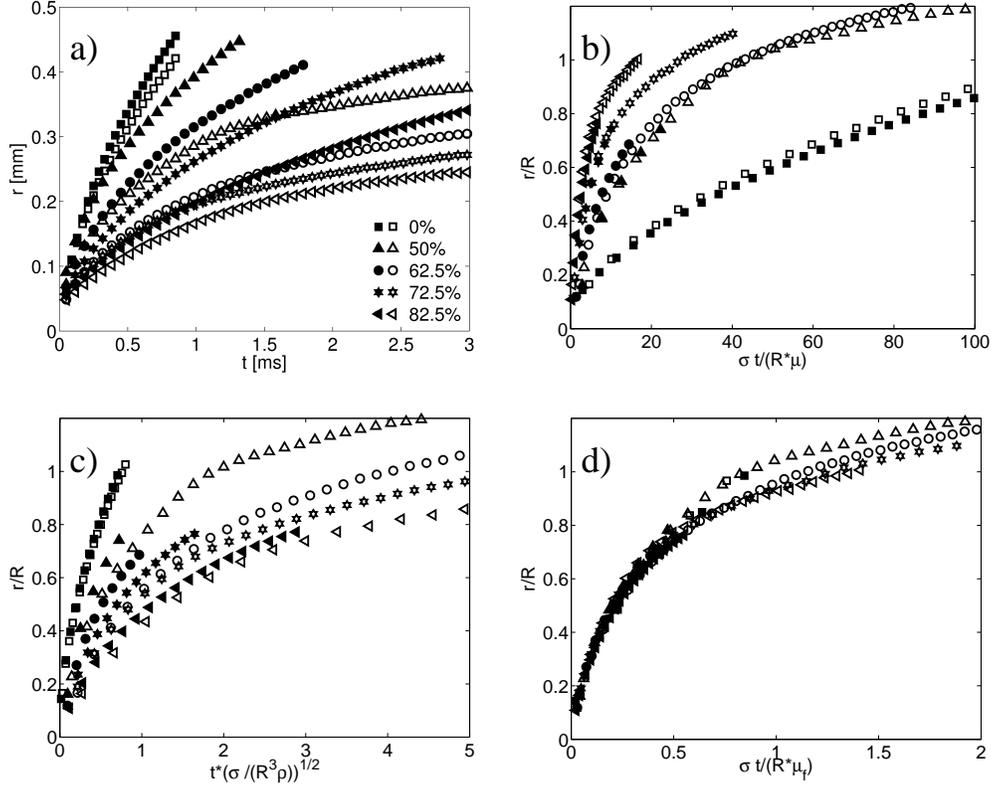}}
%\begin{minipage}[b]{0.475\linewidth}
%\label{fig:1a}\centering
%{\includegraphics[width=8cm]{fig1a_2Gimp}}
%\end{minipage}
%\begin{minipage}[b]{0.475\linewidth}
%\label{fig:1b}\centering
%{\includegraphics[width=8cm]{ViscousSiO2}}
%\end{minipage}
%\begin{minipage}[b]{0.475\linewidth}
%\label{fig:1c}\centering
%{\includegraphics[width=8cm]{InertiaSiO2}}
%\end{minipage}
%\begin{minipage}[b]{0.475\linewidth}
%\label{fig:1d}\centering
%{\includegraphics[width=8cm]{FrictionSiO2}}
%\end{minipage}
\caption{The spreading radius in time on an oxidized Si-wafer for two drop sizes $R\approx(0.3\pm0.02)$mm (hollow markers) and $R\approx(0.5\pm0.02)$mm (filled markers) for different mass fractions glycerin as indicated in the legend. (a) Dimensional units. (b) Viscous scaling. (c) Inertial scaling. (d) Contact line friction scaling.}
\label{fig:dissipation}
\end{figure}

Fig.2a shows how the radial position of the contact line evolves in time for drops with different initial radii and for different viscosities on the oxidized Si-wafer. The markers represent the mean value after several realizations of the experiments (minimum of four) and the data set has been reduced for clarity. One observation to be made in fig.2a is that the viscosity as well as the drop size influences the spreading. 

Fig 2b shows the same data, with the contact line radius scaled with initial drop radius $R$ and the time scale with a viscous capillary speed $\sigma/\mu$. The capillary speed $\sigma/\mu$ is 73m/s for water and 0.75 m/s for 85$\%$ glycerin-water. However, as is evident from fig 2b, this scaling fails to collapse the data, so the viscous contribution does not seem to be the limiting factor in this situation. An alternative would be an inertial scaling of time based on an inertial capillary velocity scale $\sqrt{\sigma/(\rho R)}$, as shown in fig 2c. As is evident here, this scaling does not capture the essential dynamics either, and we conclude that neither inertia or bulk viscosity is the limiting factor for spreading in our experiments. 

The remaining possibility is a capillary velocity based on the contact line friction discussed above and quantified in table 1. A representative velocity in this case can be found either from equation \ref{eq:cor} or from dimensional analysis to be $u^*=\sigma/\mu_f$. Introducing the values for $\sigma$ and $\mu_f$ from table 1 gives a speed of $u^*_{0\%}\sim4.8m/s$ for water and $u^*_{85\%}\sim0.6m/s$ for $85\%$ glycerin. By scaling time with $R/u^*$, we do obtain a collapse of data, for the entire range of viscosities and drop sizes, see fig.2d. The scattered dimensional plot represented in fig.2a is reduced to nearly a single spreading curve. Fig. 2 shows only results for the $SiO_2$ surface, but similar results are also obtained for the other solid surface coatings.

$\mu_f$ is determined by adjusting it in simulations so that the mean spreading radius agrees with that of several experiments performed using the drop radius 0.5mm. It should be noted that the adjustment of this single parameter achieves excellent agreement for the entire drop shape, over the whole spreading event. We have also varied the drop size in additional experiments, which has a significant influence on the spreading radius (see fig.2a). As shown in fig 2d, the data for both drop sizes collapse excellently when using a scaling of time according to $\sigma t/(R*\mu_f)$. The value of $\mu_f$ is thus independent of drop size, and this indicates that it is an intrinsic material property of the surface in combination with the wetting liquid.

%We would like to point out that we have from the simulations only determined a single parameter $\mu_f$, which in principle makes the equations free of any adjustable parameters. By extracting the value of $\mu_f$, the complete dynamics in the experiment is captured in the simulation. $\mu_f$ is measured from simulation by comparing with the mean spreading radius of several experiments with the same drop radius ($0.5mm$). We also change the drop radius in additional experiments, which has a significant influence on the spreading radius (see fig.2a). The drop radius enters in the scaling with $\mu_f$, where the collapse of data makes us believe $\mu_f$ to be a material property for the different three-phase systems.  

Fig.3a shows the non-dimensional collapse of data for the three surface coatings for different drop sizes and viscosities. By representing the dimensionless curves in fig.3a in logarithmic axis, we observe that the radii follow the same slope independent of the solid surface at the early stage of the partial wetting, see fig.3b. This is indicating that the governing physical mechanism is indeed the same for the different solid surfaces. From fig.3b it is clear that the spreading radius evolves as $\frac{r}{R}\sim(\frac{\sigma t}{R \mu_f})^{{\frac{1}{2}}}$. A similar relationship is expected in a diffusion process, where in this context $\sigma R/\mu_f$ would represent a diffusion coefficient. This could indicate that a diffusive process is taking place at the contact line, which was suggested by \cite{Ren:2007qc} from rapid wetting simulations using molecular dynamics. However, a detailed study at the nanoscale would be needed to verify this. In the first stage of the spreading, for non--dimensional time $<1$, the experiments cannot be fully captured by the hydrodynamic theory through Tanner's law $r=R(\frac{\sigma t}{\mu R})^{{\frac{1}{10}}}$ or by the molecular kinetic theory that predicts $r\sim t^{\frac{1}{7}}$ \cite{Ruijter:1999kx, De-Coninck:2001uq}. We have for clarity inserted the slope predicted from Tanner's law in fig.3b.

In fig.3b a distinct transition between the 1/2 slope and a much more gradual slope ($\sim$ 1/10) is observed around non-dimensional time $1$. This might be an indication of the transition between contact line friction dominated spreading and another slower spreading regime. We assume here that the second regime is viscously dominated spreading given by Tanner's law and makes this equal to the contact line friction dominated spreading $r=R\sqrt{\frac{\sigma t}{\mu_f R}}$ a distinct transition time ($t_t$) between the two regimes is obtained. In dimensional scales this becomes $t_t=\frac{R \mu_f}{\sigma} (\frac{\mu_f}{\mu})^{{\frac{1}{4}}}$ or in non-dimensional time ($\tau$), $\tau=\frac{t_t\sigma}{R \mu_f}= (\frac{\mu_f}{\mu})^{{\frac{1}{4}}}$. Introducing the material properties in the expression for $\tau$ we notice that a physically reasonable transition time is obtained and in very good agreement with the experimental results presented in fig.3. For example, the dimensionless transition time for water and $85\%$ glycerin is on the oxide surface found to be $\tau_{0\%}=3.5$ and $\tau_{85\%}=1.05$, respectively.

An analytical function can be derived for the contact line velocity ($\hat u_{cl}$), based on the boundary condition given in eq.\ref{bc}, if the equilibrium profile for the concentration across the interface is introduced and some algebra is performed \cite{Yue:relax2011},
\begin{equation}
\hat u_{cl}=\frac{\sigma}{\mu_f}\frac{\cos(\theta_e)-\cos(\theta)}{\sin(\theta)}
\label{eq:cor}
\end{equation}
where $\theta$ is the dynamic or apparent contact angle. Eq. \ref{eq:cor} is different from other expressions for the contact line velocity previously reported in the literature \cite {degennes:1985} by that it is divided by $\sin(\theta)$ which makes the expression diverge at angles $0^{\circ}$ and $180^{\circ}$. This function is assumed to only be valid when the local dissipation at the contact line dominates. At these extrema, other mechanisms such as inertia or bulk viscous friction are expected to regularize the solution. $\sin(\theta)$ gives a non-negligible contribution to the function and introduces an additional non-linearity.

In fig.3b it is clear that the spreading radius evolves as a function $r\sim R(\frac{\sigma t}{R \mu_f})^{{\frac{1}{2}}}$. By taking the time derivative of this expression we find that the contact line speed should be proportional to $\sim t^{-\frac{1}{2}}$. To evaluate the analytical expression for the contact line velocity given in eq.(2), we use the experimental data for the dynamic contact angle for the data presented in fig.2a for the different viscosities and drop sizes, as they evolve on the oxidized wafer. We define the dynamic contact angle between the tangent along the contoured interface (interpolated at a fixed height of 7 pixels from the wall) and the solid substrate, on the liquid side. The dynamic contact angle measurements are found to be fairly insensitive to changes in interpolation height, as long as this height is chosen to be less than the local radius of curvature at the contact line. 

Fig.4 shows that the expression given in eq.2 indeed gives a slope for the contact line speed as $\hat u_{cl}\sim R\sqrt{\frac{\sigma}{R \mu_f}} t^{-\frac{1}{2}}$. This indicates that $\frac{\cos(\theta_e)-\cos(\theta)}{\sin(\theta)}\sim R/r$, which from eq.2 recovers the experimentally observed behavior presented in fig.2d and fig.3. The inset in fig.4 shows the predicted contact line speed using the linearized function from molecular kinetic theory $\hat u_{MKT}=(\sigma/\mu_f)\cdot(\cos(\theta_e)-\cos(\theta))$ \cite{Ruijter:1999kx}. Since we are interested in the slope for the contact line speed in time, we assume $\mu_f$ to be the same in $\hat u_{MKT}$ as reported in table I. One clear observation to make from the inset in fig.4 is that at non--dimensional time $<2.4$ the slope for the contact line speed predicted from molecular kinetic theory $u_{MKT}=(\mu_f/\sigma)\hat u_{MKT}$ does not agree with the experimental observation in fig.3.\\
\\
In summary we have shown that spreading experiments and simulations for a wide range of viscosities, on substrates with very different wetting properties, all exhibit a universal spreading behavior if contact line friction dominates the spreading. An expression for the contact line radius is proposed for this spreading regime as $r\sim R(\frac{\sigma t}{\mu_f R})^{\frac{1}{2}}$. The analytical contact line velocity from phase field theory, where the dynamic contact angle is primary input, predicts the same slope for the spreading as found directly in experiments. A criterion is proposed to determine the dominance of contact line friction or viscosity in spreading, which is found in good agreement with the experiments. We hope that these results can help rationalize spreading phenomena that falls beyond classical hydrodynamic theory, and gives a phenomenological explanation for such physics. 

\begin{figure}[h!]
{\includegraphics[width=0.60\linewidth]{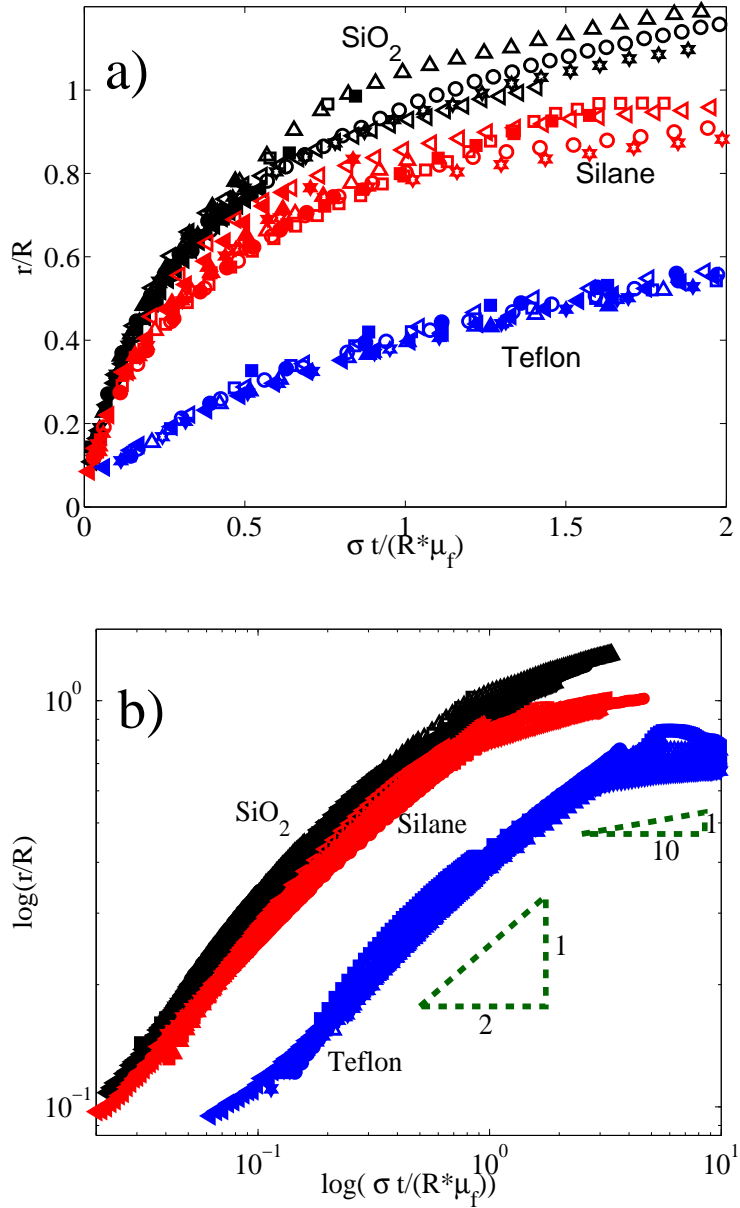}}
%\begin{minipage}[b]{0.475\linewidth}
%\label{subfig:dis-sa}\centering
%{\includegraphics[width=8cm]{ScaledFricAll}}
%\end{minipage}
%\begin{minipage}[b]{0.475\linewidth}
%\label{subfig:dis-sb}\centering
%{\includegraphics[width=8cm]{FIG2}}
%\end{minipage}
\caption{Non-dimensional spreading radius based on a contact line friction scaling on the different substrates (oxide (back), silane (red), teflon (blue)). Hollow markers denote ($R=0.3mm$) and filled markers ($R=0.5mm$). (a) Linear axis. (b) Logarithmic axis.}
\label{fig:rvst}
\end{figure}

\begin{figure}[ht!]
\centering
\includegraphics[width=0.60\linewidth,angle=0]{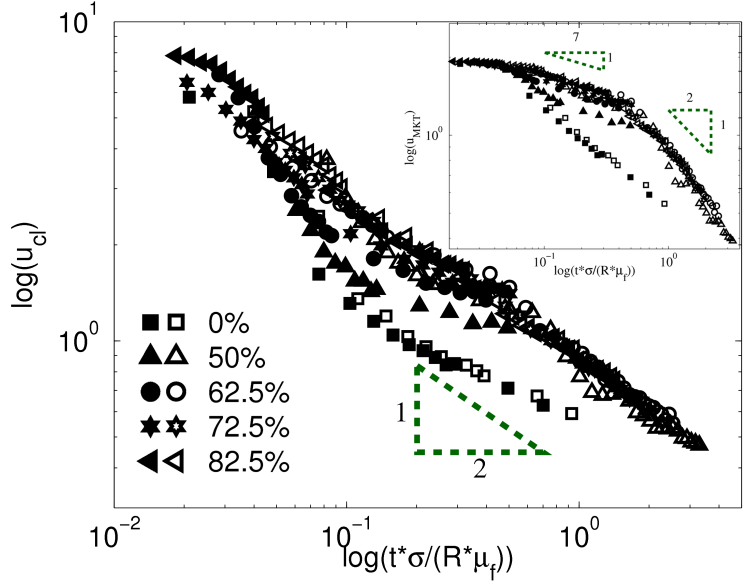}
\caption{The main figure shows the dimensionless contact line velocity function from phase field theory $u_{cl}= \frac{\mu_f}{\sigma}\hat u_{cl}=(\cos(\theta_e)-\cos(\theta))/\sin(\theta)$. Inset shows the velocity predicted from the linearized molecular kinetic theory $u_{MKT}=(\mu_f/\sigma)\hat u_{MKT}=\cos(\theta_e)-\cos(\theta)$. The input in these two functions are the experimentally measured dynamic contact angle $\theta$ for two different drop sizes on the oxidized Si-wafer. The mass fraction glycerin is indicated in the legend. Hollow markers denote small ($R=0.3mm$) and filled markers large drops ($R=0.5mm$).}
\label{fig:uclvst}
\end{figure}

\clearpage
\bibliographystyle{apsr}

\end{document}